# Trusted–HB: a low-cost version of HB$^+$ secure against Man-in-The-Middle attacks

Julien Bringer and Hervé Chabanne *

**Abstract**

Since the introduction at Crypto'05 by Juels and Weis of the protocol HB$^+$, a lightweight protocol secure against active attacks but only in a detection based-model, many works have tried to enhance its security. We propose here a new approach to achieve resistance against Man-in-The-Middle attacks. Our requirements – in terms of extra communications and hardware – are surprisingly low.

## 1 Introduction

Radio Frequency IDentification (RFID) systems are still a great challenge for researches in the field of security and privacy. One main problem is the need for ultra-lightweight cryptographic protocols.

At Crypto'05, HB$^+$, a now famous cryptographic authentication protocol very well suited for low-cost hardware implementation, was introduced by Juels and Weis [9]. It enables tags to identify themselves to the reader. HB$^+$ is presented as an improvement of the Hopper and Blum (HB) authentication scheme [7]. The security of these protocols relies on the hardness of the computational Learning Parity with Noise (LPN) problem [1, 2, 5, 8, 13]. This protocol HB$^+$ is proved secure against active attacks, though preserving HB's advantages: mainly, requiring so few resources to run that it can be implemented with only a few gates on an RFID tag. However, at the same time, Gilbert, Robshaw and Sibert [6] describe a Man-in-The-Middle (MiTM) attack on HB$^+$ not covered by the corresponding security model.

---

*J. Bringer and H. Chabanne are with Sagem Sécurité, Eragny, France. This work was partially supported by the french ANR RNRT project T2TIT.

Since this attack, various modifications of HB$^+$ have been proposed to increase its security [3, 4, 14, 15, 16]. However, none has succeeded yet to state a formal security against MiTM attacks.

In this paper, we take a new and very natural approach. We still use the protocol HB$^+$ as an identification scheme but also a way to initiate a confidential channel to authenticate the tag in a more classical manner in a second phase.

## 2  HB$^+$ protocol

Following [7], the HB$^+$ protocol security is based on the Learning Parity with Noise (LPN) problem. Note that several algorithms [1, 2, 8] are known to solve this problem and the recents proposal of [5, 13] are among the most efficient.

**LPN Problem.** Let $A$ a random $q \times k$ binary matrix, $\vec{x}$ a random $k$-bit vector, $0 < \eta < 1/2$ a noise parameter and $\vec{\nu}$ a random $q$-bit vector of weight $\text{wt}_H(\vec{\nu}) \leq \eta q$. Given $A$, $\eta$ and $\vec{z} = A.^t\vec{x} \oplus \vec{\nu}$, find a $k$-bit vector $\vec{x'}$ such that $\text{wt}_H(A.^t\vec{x'} \oplus \vec{z}) \leq \eta q$.

The HB$^+$ protocol is made of $r$ successive iterations of a round – as described in Fig. 1 where the two $k$-bit vectors $\vec{x}$ and $\vec{y}$ are secret keys shared by the Tag and the Reader. The Tag is successfully authenticated if the check fails at most $u \times r$ times for a given threshold $u$. Moreover, the Reader does not need to know a priori which tags and secrets are involved for the protocol to work. Eventually, [13] highlights that the sizes of $\vec{x}$ and $\vec{a}$ may differ from the one of $\vec{y}$ and $\vec{b}$ as the first ones only need to be 80-bit long to avoid guesses whereas the second ones are used to rely on the LPN problem.

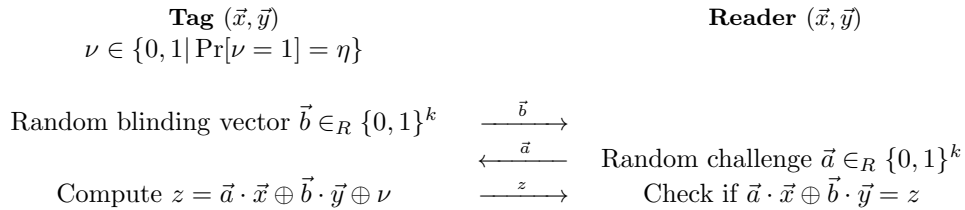

Figure 1: One round of HB$^+$

In [9], Juels and Weis prove that the protocol is secure against passive and active attacks in their security model, thanks to the difficulty of the LPN problem. Unfortunately, their model do not take into account the extra information given by the result (positive or negative) of an authentication and this is exploited during the attack introduced in [6].

The attack of [6] is a linear-time MiTM attack where an adversary located between the Reader and the Tag is able to corrupt the challenge at every round. The adversary chooses a vector $\vec{\delta}$ in $\{0,1\}^k$ and when a challenge $\vec{a}$ is sent by the Reader, he intercepts the challenge and makes a switch to $a \vec{\oplus} \delta$. Hence, at the end of the round, the Reader will receive $\tilde{z} = (a \vec{\oplus} \delta) \cdot \vec{x} \oplus \vec{b} \cdot \vec{y} \oplus \nu$ from the Tag. This is repeated along almost all the rounds in order to deduce information from the result of the authentication. If the authentication succeeds (resp. fails), we have $\vec{\delta} \cdot \vec{x} = 0$ (resp. $\vec{\delta} \cdot \vec{x} = 1$) with a high probability. So one can recover $\vec{x}$ "bit after bit" by varying $\vec{\delta}$ progressively.

The security of $HB^+$ has also been extensively analyzed to extend the protocol to parallel and concurrent executions in [10] and to explore further the large error case ($1/4 < \eta < 1/2$) in [11].

## 3 Our proposal

### 3.1 Preliminary definitions

In order to resist to Man-in-The-Middle attacks, a natural idea is to send a proof of integrity of the different parameters to the Reader. The problem is to find a lightweight algorithm to achieve this. Particularly, classical MAC algorithms, obtained from cryptographic block ciphers or cryptographic one-way hash functions seem too heavy in our case.

Interestingly, we can rely on more traditional hashing techniques following the work of Carter and Wegman [18] and specifically on the very simple construction proposed by Krawczyk [12]. Krawczyk uses in [12] a family $H$ of linear *hash* functions which map $\{0,1\}^m$ to $\{0,1\}^n$ in a balanced way following the next definition.

**Definition 1** *A family $H$ of hash functions is called $\epsilon$-balanced (or $\epsilon$-almost universal) if*

$$\forall \vec{x} \in \{0,1\}^m, \vec{x} \neq 0, \vec{c} \in \{0,1\}^n, \Pr[h \in H, h(\vec{x}) = \vec{c}] \leq \epsilon.$$

Now we suppose that the parties share a common key which consists of the choice of a particular function $h \in H$ and a random pad $\vec{e}$ of length $n$ then the message authentication of a message $\vec{x}$ is computed as $\vec{t} = h(\vec{x}) \oplus \vec{e}$.

Here, an adversary will succeed in breaking the authentication if he finds $\vec{x'}$ and $\vec{t'}$ such that $\vec{t'} = h(\vec{x'}) \oplus \vec{e}$. With respect to the simplicity of this construction, it is important that an adversary does not learn which $h$ or $\vec{e}$ is involved. If $H$ is a family of linear hash functions and if $H$ is $\epsilon$-balanced then, it is proved in [12] that the probability of success of an adversary is lower than $\epsilon$; the scheme is then said $\epsilon$-secure. This clearly emphasizes the interest of this construction.

Following the principle of a one-time pad, the same $h$ can be reused but $\vec{e}$ must be different each time, i.e. it is $\epsilon$-secure against any adversary (unconditionaly) only if $\vec{e}$ is a random pad.

**LFSR-based Toeplitz construction.** To construct such hash families, an efficient solution is provided in [12]. The author simplifies the multiplication with a boolean matrix by restricting it to specific Toeplitz matrices which can be described by a LFSR. Let the LFSR represented by its feedback polynomial $P$, an irreducible polynomial over $\mathbb{F}_2$ of degree $n$, and an initial state $\vec{s} = (s_0, \ldots, s_{n-1}) \neq 0$, then $h_{P,\vec{s}} \in H$ is defined by the linear combinations $h_{P,s}(\vec{x}) = \bigoplus_{j=0}^{m-1} x_j . \vec{S_j}$ where $\vec{S_j}$ is the $j$-th state of the LFSR (i.e. $\vec{S_0} = \vec{s}$).

Following [12], this family $H$ is then $\epsilon$-balanced for at least a $\epsilon \leq \frac{m}{2^n-1}$. Moreover, a hash $h_{P,\vec{s}}$ is easily implemented in hardware and a second advantage is that the message authentication can be computed progressively with an accumulator register which is updated after each message bit: the implementation does not depend on the size $m$ of $\vec{x}$.

### 3.2 Description

We describe here the improved version we propose for HB$^+$ to thwart Man-in-The-Middle attacks.

We now suppose that the Tag and the Reader share a key $(\vec{x}, \vec{y}, h)$ with $\vec{x} \in \{0,1\}^{k_1}, \vec{y} \in \{0,1\}^{k_2}$ and $h \in H$ for $H$ a linear and $\epsilon$-balanced hash family. The beginning stays unchanged, $r$ rounds of HB$^+$ protocol are executed (see Fig. 1), i.e. for $i$ from $0$ to $r-1$:

- $\vec{b_i} \in_R \{0,1\}^{k_2}$ is sent to the Reader;
- $\vec{a_i} \in_R \{0,1\}^{k_1}$ is sent to the Tag;

- $\nu_i \in \{0, 1\} | \Pr[\nu = 1] = \eta\}$ is taken;
- $z_i = \vec{a_i} \cdot \vec{x} \oplus \vec{b_i} \cdot \vec{y} \oplus \nu_i$ is sent to the Reader;
- the Reader checks whether if $z_i = \vec{a_i} \cdot \vec{x} \oplus \vec{b_i} \cdot \vec{y}$.

Thereafter, if the number of incorrect checks is lower than the threshold $u \times r$, the Reader waits for a last message to authenticate the Tag. This first phase – corresponding to an execution of the HB$^+$ protocol – is interpreted as a way to recover among a set of registered secrets $\{(\vec{x^j}, \vec{y^j}, h^j)\}_j$ which $(\vec{x}, \vec{y})$ has been used. Once the correct $(\vec{x}, \vec{y})$ is found, the Tag will authenticate itself with the associated function $h$.

After the $r$ rounds of this first phase, the second phase is the following.

1. Starting with a noise $\vec{\nu} = (\nu_0, \ldots, \nu_{r-1})$, the Tag computes $\vec{e} = E(\vec{\nu}) \in \{0,1\}^n$ and sends

$$\vec{t} = h\left(\left(\vec{a_0}, \vec{b_0}, z_0, \ldots, \vec{a_{r-1}}, \vec{b_{r-1}}, z_{r-1}\right)\right) \oplus \vec{e}$$

   to the Reader, following the principles of [12] recalled in Sec. 3.1.

2. For all $i \in \{0, \ldots, r-1\}$, the Reader recovers $\nu_i = z_i \oplus \vec{a_i} \cdot \vec{x} \oplus \vec{b_i} \cdot \vec{y}$, computes $\vec{e} = E(\vec{\nu}) \in \{0,1\}^n$ and it checks the validity of the received tag $\vec{t}$ with respect to the received words $\vec{a_0}, \vec{b_0}, z_0, \ldots, \vec{a_{r-1}}, \vec{b_{r-1}}, z_{r-1}$.

Here $E$ maps a $\eta$-biased vector in $\{0,1\}^r$ to a quasi-random vector of $\{0,1\}^n$ (cf. section 3.4) and $h$ is defined over $\{0,1\}^m$ with $m = r.(k_1 + k_2 + 1)$.

If the verification succeeds then the authentication is done.

Informally, the original HB$^+$ protocol helps to identify the Tag meanwhile it enables to transmit a pseudo-random pad $\vec{e}$ to the Reader. This information enables us to construct a final message authentication which aim is to prove the integrity of the communications.

### 3.3 Security arguments

First, the protocol is obviously correct as the last verification is straightforward when there is no perturbation of the communications. Secondly, with a good pseudo-random function $E$ the last iteration would bring no useful information for solving the LPN problem with secrets $(\vec{x}, \vec{y})$ so it seems to inherit the security of HB$^+$ against passive and active (not MiTM) attacks. Moreover, we have:

**Theorem 1** *If the message authentication scheme induced by the hash family $H$ is $\epsilon$-secure and if the output of $E$ is random and unknown, then any MiTM adversary has a probability of success of at most $\epsilon$.*

*Sketch of the proof.* Indeed, an adversary has a probability at most $\epsilon$ of being authenticated with modified communications. Suppose that the Tag has received altered challenges $\vec{a'_0}, \ldots, \vec{a'_{r-1}}$ and that the Reader received modified answers $\vec{b'_0}, \ldots, \vec{b'_{r-1}}, z'_0 \ldots, z'_{r-1}$ and a message authentication tag $\vec{t'}$. To be valid, $\vec{t'}$ must be equal to

$$h\left(\left(\vec{a_0}, \vec{b'_0}, z'_0, \ldots, \vec{a_{r-1}}, \vec{b'_{r-1}}, z'_{r-1}\right)\right) \oplus E(\vec{\nu'})$$

with $\nu'_i = z'_i \oplus \vec{a_i} \cdot \vec{x} \oplus \vec{b'_i} \cdot \vec{y}$. If $\vec{\nu'}$ is unknown from the adversary, then it happens only with a probability lower than $\epsilon$ thanks to [12]. □

Note that the knowledge of $\vec{\nu'}$ is conditioned by the difficulty to retrieve $\vec{x}$ and $\vec{y}$ from the communications.

### 3.4 Implementation

Here we only add one iteration to the $r$ iterations of HB$^+$. Moreover as mentioned in section 3.1, a LFSR-based Toeplitz hashing is easy to embed in hardware circuits. It is still the case even with an important number of rounds: we can take advantage of the construction to compute progressively the last authentication message $\vec{t}$ round after round thanks to an accumulator which is updated input's bit by input's bit. Thus the computation cost is low.

The main question remains on the function $E$ which must ensure a good randomness of its output with the biased vector $\vec{\nu}$ as input.

We might use a randomness extractor to implement $E$. For instance, if we assume that the bit of $\vec{\nu}$ are independent and identically distributed (as it is for the analysis of the LPN solving algorithms such as [13]), the von Neumann procedure [17] outputs a sequence of statistically independent and equiprobable bits. On an input source $x_1, \ldots, x_N$, it considers pairs $(x_{2i+1}, x_{2i+2})$ and outputs $x_{2i+1}$ if they differ, nothing otherwise. For a bias $\eta$, from a source of length $N$, the output has a mean length of $N \times \eta(1-\eta)$.

**Example of parameters.** Following [13], we choose for the underlying HB$^+$ protocol $\eta = 0.25$, $k_1 = 80$ and $k_2 = 512$ to ensure 80 bits security with respect to the best known algorithm to solve instances of the LPN problem. In this case, with a threshold $u = 0.348$ and $r = 1164$ rounds, the probability to reject a genuine tag will be about $2^{-40}$ and the probability of authentication with random guesses will be close to $2^{-80}$.

The size $m = r.(k_1 + k_2 + 1)$ of the final message to authenticate is then sufficiently large and if we use the von Neumann extractor it leads to a mean output's length 218 with a standard deviation about 13. In practice, we restrict ourselves to the first $n$ bits with $n = 101$. With a LFSR-based Toeplitz hash family of [12], it enables us to achieve an $\epsilon$-secure message authentication algorithm with $\epsilon \leq 2^{-80}$. Note that the probability to extract less than 101 bits in this situation is lower than $2^{-72}$ so it is unlikely to happen (if it happens, the authentication process could restart).

## 4  Conclusion

Traditional remedies to the MiTM problems of HB$^+$ work fine. The addition of a cryptographic check of the communications prevents an adversary to modify the exchanges between a Tag and its Reader. The reuse of the techniques of Krawczyk [12] for enforcing integrity is here determinant as this enables us to propose a solution which is still suitable for low-cost Tags.